# X-ray diffraction study of structure and molecular correlations in liquid 1,3,5–triphenylbenzene at 473 K

Henryk Drozdowski[a,*], Małgorzata Śliwińska-Bartkowiak[a]

[a] *Faculty of Physics, Adam Mickiewicz University, Umultowska 85, 61-614 Poznań, Poland*

**A B S T R A C T**

The structure of 1,3,5–triphenylbenzene $C_6H_3(C_6H_5)_3$ at 473 K was investigated using the X-ray diffraction method. The measurements of scattered radiation intensity were performed in a wide range of wave vector ($S_{min} = 0.925$ Å$^{-1}$ to $S_{max} = 14.311$ Å$^{-1}$). For the first time the theoretically predicted model of the structure 1,3,5–triphenylbenzene was experimentally confirmed. The model of short-range arrangement of the molecules was proposed. The determined mean, smallest mutual distances between molecules of liquid studied are: $\bar{r}_1 = 4.30$ Å, $\bar{r}_2 = 5.30$ Å, $\bar{r}_3 = 5.40$ Å. The most probable value of the packing coefficient of the molecules was found to be $k = 0.53$. This value falls in the range of $k$ values permissible for liquid phase. The liquid studied is a major aromatic compound. The study of structure of 1,3,5–triphenylbenzene may be helpful in the explanation of the mechanism of intermolecular interactions in synthetic polymers.

## 1. Introduction

An elementary cell of a crystalline 1,3,5–triphenylbenzene belonging to the orthorhombic system /$a = 7.610(2), b = 19.765(2), c = 11.258(2)$/ Å, space group $Pna2_1$) contains 4 molecules [1]. However, to the best of our knowledge, 1,3,5–triphenylbenzene has not been studied in the liquid phase by diffraction methods. The present article deals with the structural study of 1,3,5–triphenylbenzene at 473 K. The structure of liquid 1,3,5–triphenylbenzene was studied by diffraction of strictly monochromatic X-ray radiation. The angular distributions of the X-ray scattered intensity were measured for the angles $6° \leq 2\Theta \leq 120°$ at every 0.2°, where $\Theta$ is the Bragg angle. The measurements were performed using the transmission technique. The method is described in details in [2,3].

A value of the packing coefficient was proposed to be a criterion of correctness of the liquid structure model determined. For the proposed model of molecule 1,3,5–triphenylbenzene, the theoretical function of the reduced intensity was calculated and compared to the experimental one.

The differential radial-distribution function of electron density was numerically calculated using Fourier analysis. The mean distances between the neighbouring molecules and the mean coordination numbers were found. From the known volume of the first coordination sphere as well as the specific volume of the molecule and taking into regard the structural model of intermolecular interactions.

---
[*] Corresponding author. Tel.: +48 61829 5029, fax: +48 61829 5155.
E-mail address: riemann@amu.edu.pl (H. Drozdowski).

## 2. Experimental method

X-ray scattering in liquid 1,3,5–triphenylbenzene (melting point 444 K) was measured by applying MoKα radiation, $\lambda = (0.71069 \pm 0.00012)$ Å. The radiation was monochromatized by reflection from the (002) planes of flat graphite with the angle of monochromatization of $\alpha = 6°00'$. The angular distributions of the X-ray scattered intensity were measured for the angles $6° \leq 2\Theta \leq 120°$ at every $0.2°$, where $2\Theta$ is the scattering angle.

The measurements were made by the transmission method at the beam incidence and diffraction symmetric with respect to the flat surface of the preparation studied. The movement of the counter on the goniometer circle was coupled with the rotation of the table with the sample about the vertical axis of the goniometer at the ratio 2:1. The synchronization of the two movements means that the input slit of the counter is always at the point of intersection of the focus circle with goniometer circle. The geometry of the system was verified by testing the agreement between the results measured towards increasing and decreasing angles.

An important problem was the divergence of the primary beam and that diffracted in the vertical direction. In the path of the diffracted beam there was a system of Soller slits composed of a few thin metallic plane-parallel plates whose separation determined the vertical divergence of the diffracted beam. The divergence of the primary beam in the vertical direction was restricted by a set of collimation slits. The voltage on the probe was optimised to be 1.85 kV. The deviation in intensity due to instability of the diffractometer work were of about 1% in the whole range considered.

X-ray diffraction investigation of liquid 1,3,5–triphenylbenzene was performed at 473 K. Triphenylbenzene samples of 99% purity were purchased from Aldrich – Chemie (Germany). Studies of 1,3,5–triphenylbenzene $C_{24}H_{18}$ in the liquid phase have been very difficult because of its sublimation properties. The liquid 1,3,5–triphenylbenzene was placed in an electrically heated cell [3] closed on both sides with windows made of Bengal mica of $0.015 \pm 0.001$ mm in thickness. The high temperature camera was previously used to study the structural of naphthalene derivatives [3].

## 3. Correcting the experimental data

The experimentally obtained function of the angular distribution of the scattered X-ray intensity was corrected to include the background [4], polarization [5], absorption [6] and anomalous dispersion [7]. For a flat preparation of the thickness $D$, the absorption coefficient applied in the transmission method was [6]:

$$A(2\Theta) = \frac{\mu \cdot D(\sec 2\Theta - 1)}{1 - \exp[-\mu \cdot D(\sec 2\Theta - 1)]}, \qquad (1)$$

where $\mu$ – is the linear absorption coefficient of a liquid studied, $D$ is the thickness of the sample studied and $2\Theta$ is the scattering angle.

The coefficient $\mu$ of Eq. (1) is calculated for more complex molecules from an expression for the mass absorption coefficient given by the relation:

$$\frac{\mu}{\rho} = \frac{1}{M} \sum_{i=1}^{n} \left(\frac{\mu}{\rho}\right)_i m_i, \qquad (2)$$

with $M$ denoting the molecular mass, $m_i$ the atomic mass of the $i$-th atom, and $\rho$ the macroscopic density of the liquid.

The original beam of X-ray radiation was monochromatized by reflection from a planar graphite crystal and the polarisation factor was calculated from the equation [8]:

$$P(2\Theta) = \frac{1 + \cos^2 2\Theta_m \cdot \cos^2 2\Theta}{1 + \cos^2 2\Theta_m}, \qquad (3)$$

where $2\Theta$ is the angle at which the X-ray beam was reflected from the monochromator surface. In our experiment when the $MoK_\alpha$ X-ray beam was reflected from the plane (002) the angle was $\Theta_m = 6°00'$.

The experimental values of scattered radiation intensity were corrected by the Renninger and Kaplow computer program [9] according to the scheme:

$$I = (I^{EXP} - I^{INC} - I^{MULT} - T) P A, \qquad (4)$$

where $I^{EXP}$ is the experimentally obtained intensity of scattered radiation, $I^{INC}$ – intensity of incoherent radiation, $I^{MULT}$ – intensity of multiple scattering, $T$ – apparatus background and noise of the analysing system, $P$ – polarizing factor, and $A$ – absorption factor.

Program calculating functions of radial distributions *X–Ray Structural Study of Soft Matter* it was developed in our laboratory. The scattered X-radiation was normalized to electron units [e.u.] according to the Krogh-Moe [10] and Norman [11] method. The fundamental idea of normalization is based on the finding that for large scattering angles all interference – both interatomic and intermolecular – disappears. Then, for the angular range considered, the intensity distribution of the experiment covers the theoretical curve. The calculation of the theoretical curve requires knowledge of the atomic composition of the smallest structural unit, the 1,3,5–triphenylbenzene molecule.

For a given atomic composition of a scattering structural unit, the curve of intensity of independent scattering can be obtained from tabulated atomic scattering factors [7,8]. The shape of this curve depends on a given scattering structural unit. For the liquid studied the following normalization relationship was obtained:

$$C \int_0^\infty I(S) S^2 \, dS = \sum_{j=1}^{n} \int_0^\infty f_j^2(S) S^2 \, dS, \qquad (5)$$

in which $C$ is a normalisation coefficient and $f_j(S)$ stands for atomic scattering factors expressed in electron units [12].

## 4. Determination of the structure of 1,3,5–triphenylbenzene. Mean amplitudes of vibration

To determine the conformation of the molecule we use the modified Debye equation [13]:

$$i_m(S) = \left[ \sum_{uc}^{N} \sum_{i \neq j}^{N} f_i f_j A_{ij} \frac{\sin(S \bar{r}_{ij})}{S r_{ij}} \right] \cdot \left[ \sum_{i}^{N} f_i(S) \right]^{-2}, \qquad (6)$$

In this equation $\bar{r}_{ij}$ is the distance between two atoms $i$ and $j$ (which may or may not be linked by a chemical bond), and $A_{ij}$ is an exponential term which allows for the fact that the atoms within the $N-$atomic molecule are not strictly at rest but are vibrating with respect to each other. $A_{ij}$ has the form $\exp\left[-\frac{\langle u_{ij} \rangle}{2} \cdot S^2\right]$, where $\langle u_{ij} \rangle$ is the root-mean-square variation in the distance $\bar{r}_{ij}$ between pairs of atoms. From the best values of these damping factors the average amplitudes of vibration of the different pairs at atoms may be obtained. In Eq. (6) $f_i$ and $f_j$ are the atomic scattering factors for the $i-$th and $j-$th atoms.
The sum of atomic scattering factors is determined by the expression:

$$\sum f_j (S) = 3 f_{C_6H_5} + 3 f_{CH} + 3 f_C. \qquad (7)$$

and then

$$i_m(S) = \left[ 3 f_{C_6H_5} + 3 f_{CH} + 3 f_C \right]^{-2} \cdot \left[ \sum_{j,s}^{n} \sum_{i \neq j}^{n} f_i f_j \exp\left(-\frac{\bar{u}_{ij} S^2}{2}\right) \cdot \frac{\sin(S \bar{r}_{ij})}{S \bar{r}_{ij}} \right]. \qquad (8)$$

In molecular liquids whose molecules consist of atoms of various species and atomic factors, one considers as scattering centres not the atoms as such but rather certain functions describing the electron distribution in the liquid. For different atomic species $i, j, ..., k$ one has different electron scattering functions i.e. different atomic factors $f_i, f_j, ..., f_k$ [11]. For example, the expression: $3 f_{C_6H_5}$ can be represented as follows: $3 \left[ 6 f_C + 5 f_H \right]$. The value of the atomic scattering factors are tabulated [7,8]. Computers techniques [14,15] were used to minimise the effects of experimental errors, uncertainties in the scattering factors, and termination errors.
For calculation of mean amplitudes of vibrations $\bar{u}_{ij}$ of different pairs of atoms of a liquid studied, the empirical formula of Mastryukov and Cyvin was applied [16].

The curves of reduced radiation intensity were analysed by the method of Blum and Narten [17]. The reduced intensity $i(S)$ function, defined as:

$$i(S) = i_m(S) + i_d(S), \qquad (9)$$

where $$i(S) = \frac{\bar{I}_{eu}(S)/N - \sum_{uc} f_i^2(S)}{g^2(S)} \qquad (10)$$

is the structural-sensitive part of the total coherent intensity $\bar{I}_{eu}(S)/N$ in electron units per molecule, $g(S) = \sum_{uc} f_j / \sum_{uc} Z_j$ is a sharpening factor, $\exp(-\alpha^2 S^2)$ is a convergence factor, $i_m(S)$ is the molecular structure function describing the scattering by a single molecule and $i_d(S)$ is the distinct structure function providing the information about intermolecular correlations from the experimental data.

There are two possible of the conformations of 1,3,5–triphenylbenzene, different values of the angles of twist phenyl rings (relative to the center). Farag reported the twist angles of the phenyl rings in this compound as +34, –27, and +24° [18]. The second conformation of the compound present Lin and Williams. According to these angles of rotation are phenyl rings, respectively +40.7, –37.2, +36.1° [18]. Determining the conformation was essential for the analysis of mutual positions of molecules.

The normalized function $\bar{I}(S)$ of scattered radiation intensity obtained for the liquid studied is presented in Fig. 2. The positions of the maxima on these functions were found using the Lagrange polynomials method. Using the experimental values of $\bar{I}(S)$, presented in Fig. 1 and Eq. (10), the values of $i(S)$, and the total functions of the structure were calculated.

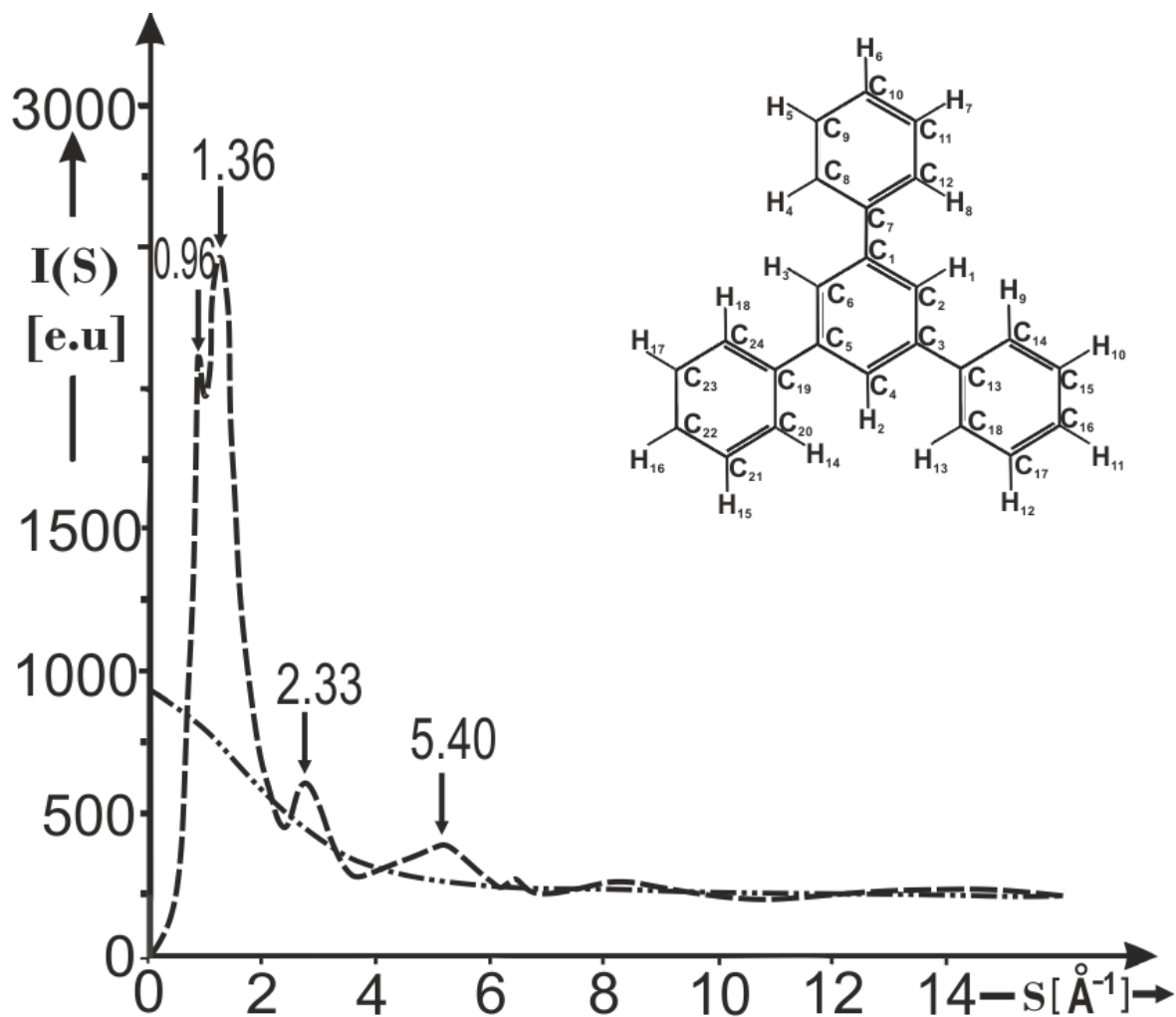

**Fig. 1.** The normalized, experimental curve of angular distribution of X-ray scattered intensity in liquid 1,3,5–triphenylbenzene using high-temperature camera.

The courses of the dependencies $i(S)$ and $i_m(S)$ for the studied liquid is shown in Fig. 2. For the properly chosen values of molecular parameters $\bar{r}_{ij}$ and thermal factor, the molecular function describing the structure of studied liquid was fitted to the experimental function of reduced intensity of scattered X-ray radiation for $S \geq 5$ Å$^{-1}$. Molecular parameters $\bar{r}_{ij}$ (Table 1) have been fitted by a testing method [13,17] assuming that $i(S) \approx i_m(S)$ for values of $S$ ($S \geq 5$ Å$^{-1}$).

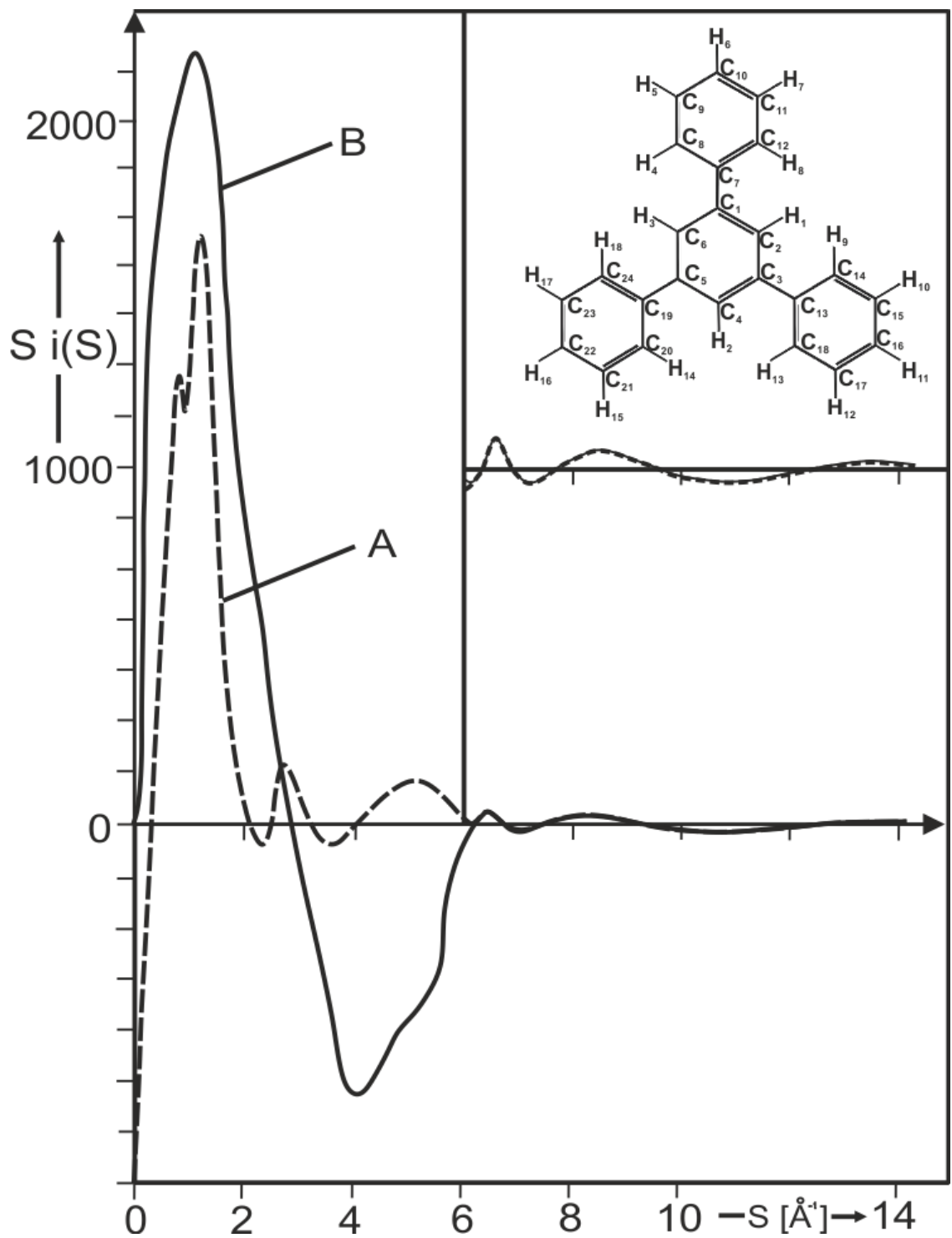

**Fig. 2.** Curve – dotted line (A), the experimental structure function $S\,i(S)\,\exp(-\alpha^2 S^2)$.
Curve – continuous line (B), the molecular structure function $S\,i_m(S)$ calculated according to Debye for 1,3,5–triphenylbenzene.

- 8 -**Table 1**
The values of parameters of 1,3,5–triphenylbenzene molecule model
applied in Debye formula Eq. (6)

| Type of intramolecular interactions | Intramolecular distances $\bar{r}_{ij}$ [ Å ] | Mean amplitude $<u_{ij}>$ [ Å ] |
|---|---|---|
| $C_2 - H_1$ | 1.10 | $0.07_7$ |
| $C_1 - C_2$ | 1.37 | $0.04_6$ |
| $C_1 - C_7$ | 1.49 | $0.03_8$ |
| $C_3 - C_{13}$ | 1.50 | $0.04_3$ |
| $C_5 \cdots C_{19}$ | 1.50 | $0.04_3$ |
| $C_1 \cdots C_3$ | 2.38 | $0.06_9$ |
| $C_1 \cdots C_4$ | 2.73 | $0.07_7$ |
| $C_1 \cdots C_9$ | 3.68 | $0.06_8$ |
| $C_1 \cdots C_{10}$ | 4.20 | $0.06_8$ |
| $C_1 \cdots C_{11}$ | 3.80 | $0.09_5$ |
| $C_3 \cdots C_{15}$ | 3.80 | $0.09_5$ |
| $C_3 \cdots C_{16}$ | 4.20 | $0.06_7$ |
| $C_3 \cdots C_{17}$ | 3.80 | $0.09_5$ |
| $C_5 \cdots C_{21}$ | 3.70 | $0.06_8$ |
| $C_5 \cdots C_{22}$ | 4.20 | $0.06_7$ |
| $C_5 \cdots C_{23}$ | 3.70 | $0.06_8$ |

Determining the conformation 1,3,5–triphenylbenzene turned out to be a very difficult task. The model assumes a conformational analysis of the test molecule (Fig. 3) on the basis of crystal X-ray data for organic compounds (inter-atomic distances, angles of rotation, the radius of the atomic van der Waals) and the selected test values of internuclear distances $\bar{r}_{ij}$. Experimental distributions of scattered X-radiation intensity were compared with theoretical results (Fig. 2) predicted for a proposed model of 1,3,5–triphenylbenzene molecule (Fig. 3). X-ray structural analysis of the liquid 1,3,5–triphenylbenzene confirmed the model of molecule proposed by Lin and Williams for crystal [18].

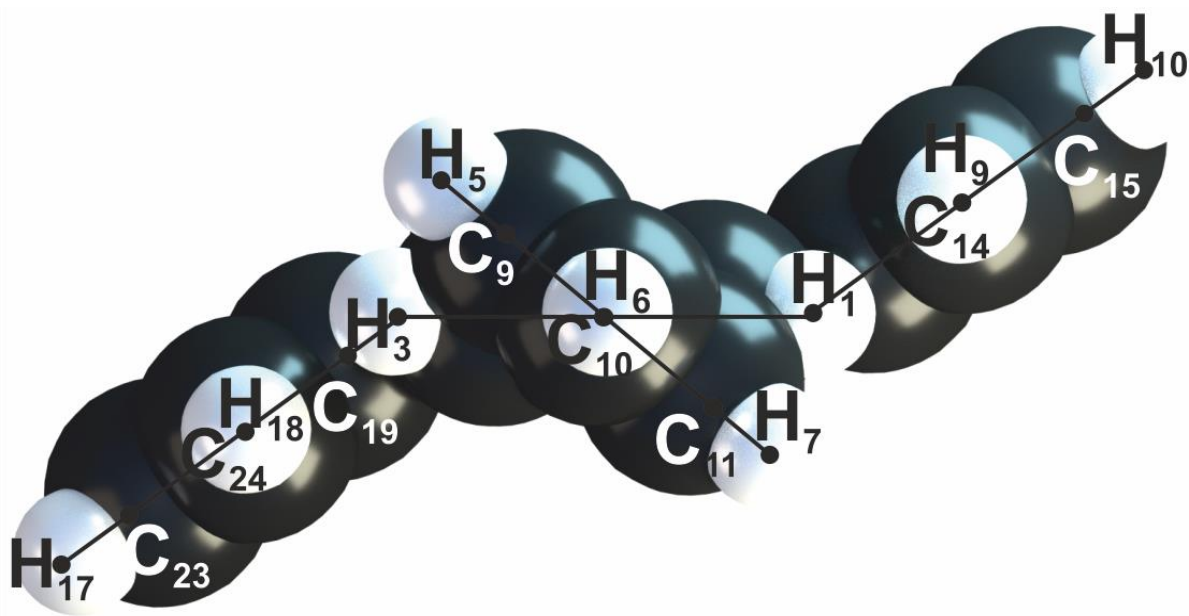

**Fig. 3.** A model of 1,3,5–triphenylbenzene $C_6H_3(C_6H_5)_3$ molecule structure: the angles of rotation of phenyl rings are as follows: $+40.7°, -37.2°, +36.1°$.

Distances characteristic for the 1,3,5–triphenylbenzene molecule are summarized in Table 1. The determined intramolecular distances compared to the corresponding results obtained by Lin and Williams are within the measurement uncertainty ($\Delta \bar{r} \in <0.01 \div 0.02> $ Å. Negative induction effect of phenyl groups in the 1,3,5–triphenylbenzene molecule, the length of the bonds between the following carbon atoms in the central benzene ring: $C_1 - C_2 \equiv C_3 - C_4 \equiv C_5 - C_6 = 1.37 \pm 0.01$ Å. Such a value for these bonds was given during the fit of the function $i(S)^{exp}$ to the function $i_m(S)$ of the wave vector $S \geq 5$ Å$^{-1}$.

## 5. Intermolecular interactions in 1,3,5–triphenylbenzene at 473 K

X-ray analysis of intermolecular interactions of liquids is based on the differential radial distribution function (DRDF) [3,13]. Analysis of the shape of the DRDF $4\pi r^2 \sum_{j,k}^{n} \overline{K}_j [\rho_k(r) - \rho_0]$ (Fig. 4) permits conclusions concerning the intermolecular interactions in 1,3,5–triphenylbenzene.
This function bring the information about the difference between the observed and the average distribution of electron density [12,19]. In liquid 1,3,5–triphenylbenzene the presence of the three coordination spheres of intermolecular ordering was established.

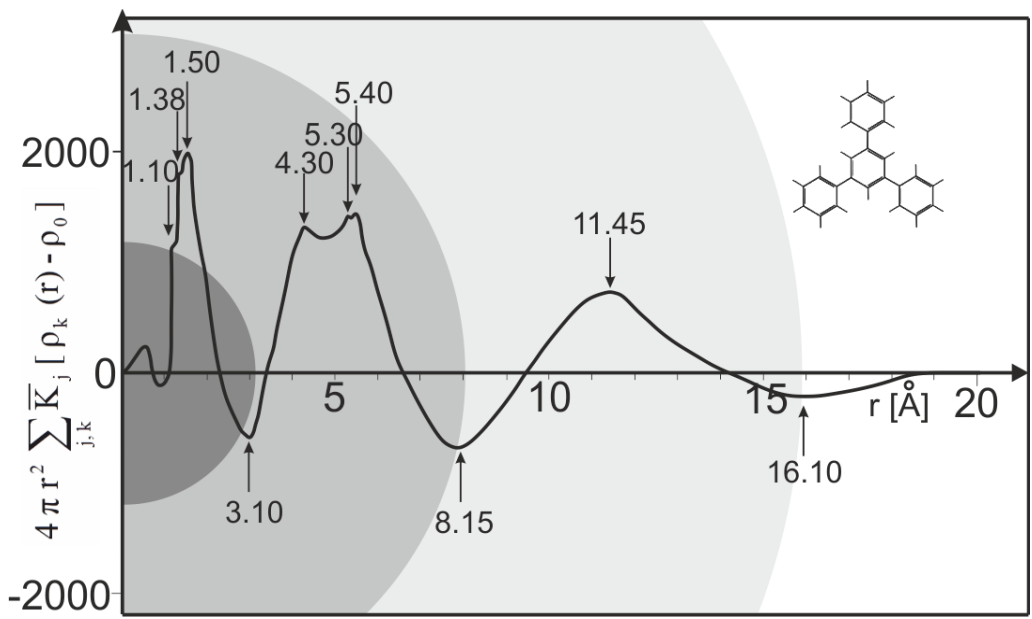

**Fig. 4.** The DRDF $4\pi r^2 \sum_{j,k}^{n} \overline{K}_j [\rho_k(r) - \rho_0]$ for liquid 1,3,5–triphenylbenzene at 473 K; the ranges coordination spheres: first sphere <3.10; 8.15> Å; the second sphere <8.15; 16.10> Å.

Subsequently, the ranges of the spheres and coordination numbers were determined. Coordination shells are delimited by minima of DRDF [19,20]. The units of the DRDF are electrons$^2$/Å. The maxima on the DRDF in the range $1 < \overline{r} \leq 3.10$ Å have been ascribed to the following pairs of atoms: $C_2 - H_1 = (1.10 \pm 0.01)$ Å, $C_1 - C_2 = (1.38 \pm 0.01)$ Å, $C_3 - C_{13} = 1.50 \pm 0.01)$ Å. The determined mean, smallest mutual distances between molecules of liquid studied are: $\overline{r}_1 = 4.30$ Å, $\overline{r}_2 = 5.30$ Å, $\overline{r}_3 = 5.40$ Å. A simple model of short-range arrangement of the molecules in liquid 1,3,5–triphenylbenzene was proposed, Fig. 5.

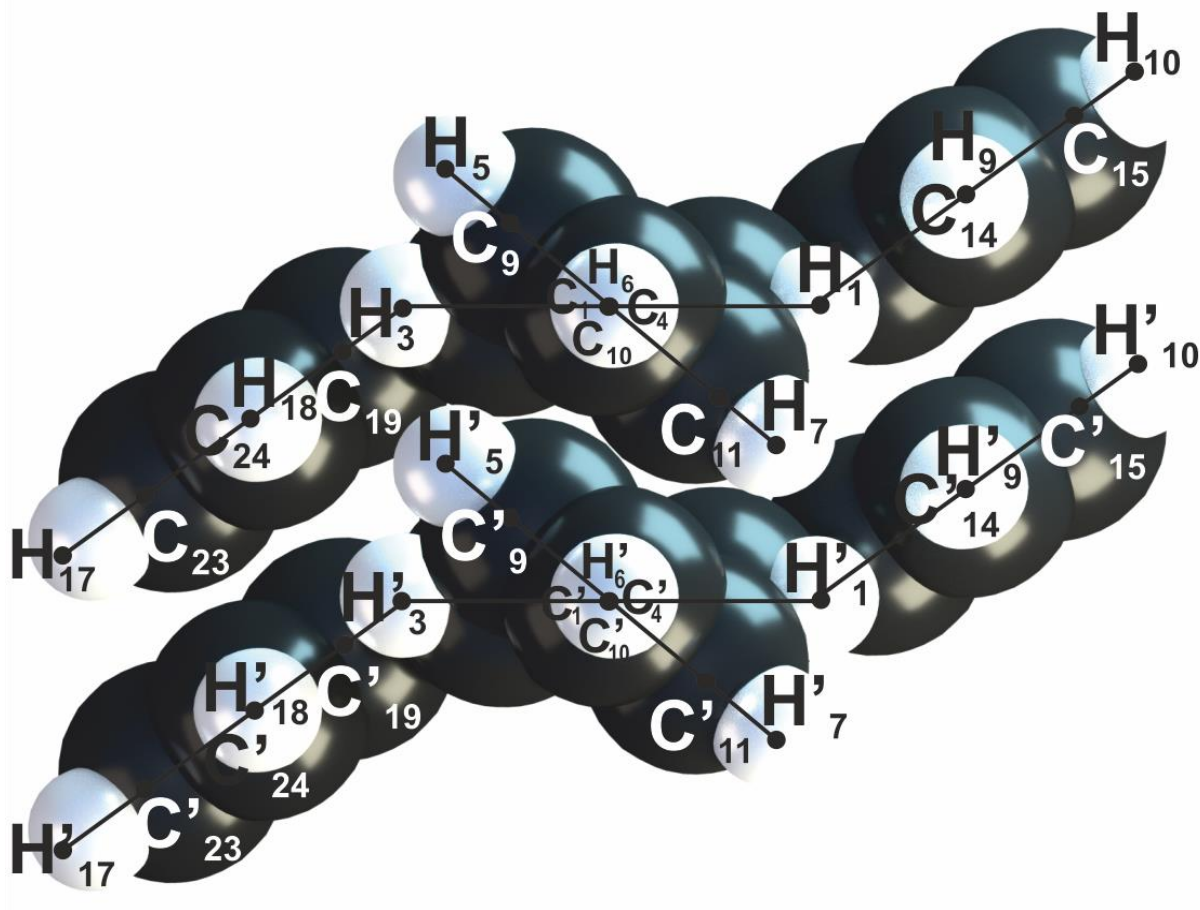

**Fig. 5.** Model of probable intermolecular interactions in liquid 1,3,5–triphenylbenzene. The arrangement of molecules corresponding to the maxima in the DRDF ($C_1 \cdots C_1' = 4.30$ Å, $C_1 \cdots C_{24}' = 5.30$ Å, $C_1 \cdots C_{14}' = 5.40$ Å).

The binary radial correlation of molecules in studied liquid is shown in Fig. 6. The most probable simple configurations of neighboring molecules in the liquid in question can be examined by fitting positions of the maxima of the DRDF to the distances between the centers of neighbors resulting from their van der Waals models [21].

By analogy with the packing coefficient defined for crystals [22] we can define the corresponding packing coefficient of molecules in liquids

$$k = \frac{V_0}{\overline{V}/x}, \qquad (11)$$

where $V_0$ is the specific volume of the molecule and $\overline{V}/x$ is the volume available for one molecule in a pseudocell [23]. The specific volume of a molecule may be found from its chemical structure, the length of bonds and van der Waals radii of atoms and functional groups. The determined value $k$ is correct only when it falls within the range $0.51 \leq k \leq 0.58$ as the values $k > 0.58$ correspond to the crystal phase and the values $k < 0.51$ to the gas phase [24].

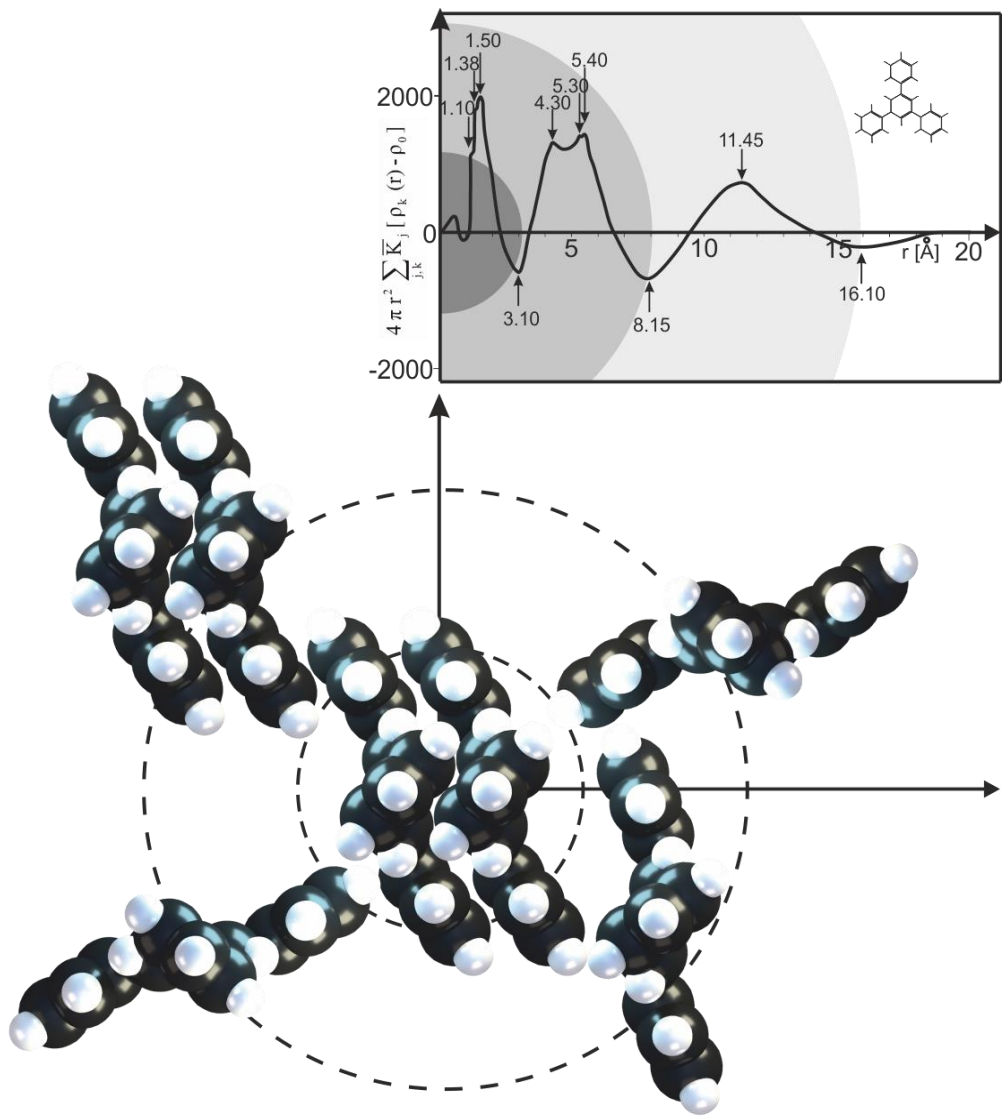

**Fig. 6.** Approximate model arrangement of molecules in 1,3,5–triphenylbenzene at 473 K proposed taking into account van der Waals interactions (dashed lines – the radius of the first and second coordination sphere); $\overline{R} \approx 5.40$ Å; $\overline{V} = 652.0$ Å$^3$, $V_0 = 315.2$ Å$^3$; $x = 1.1$; $k = 0.53$. The radius of the second coordination sphere is 11.45 Å.

The volume of the pseudocell $\overline{V}$ is found from the radius of the so-called first coordination sphere $\overline{R}$, according to the spherical relation:

$$\overline{V} = \frac{4}{3} \pi \overline{R}^3. \tag{12}$$

The mean value of the radius $\bar{R}$ is approximately determined from the position of the main maximum in X-ray scattering function:

$$\bar{R} \approx \frac{7.73}{S_{max}} - 0.3, \qquad (13)$$

where $S_{max} = 4\pi \sin\Theta_{max}/\lambda$, $2\Theta_{max}$ is the scattering angle determining the position of the main maximum of the scattered radiation intensity, and $\lambda$ is the X-ray wave length of 0.71069 Å in our experiment. The correcting factor 0.3 for molecular liquids has been found empirically [26]. From the known volume of the first coordination sphere as well as the specific volume of the molecule and taking into regard the structural model of intermolecular interactions, the most probable value of the packing coefficient of the molecules was found to be $k = 0.53$. The right coefficient value is, however, obtained only when the assumed model of intermolecular liquid structure in a given liquid is correct. If this model is not correct, the number of molecules, $x$, in one elementary pseudocell is wrongly estimated and the value of the $k$ coefficient goes beyond the limit of values admissible for liquid phase.

## 6. Conclusions

The most important results of this paper are listed below.

- New experimental data on the structure of 1,3,5–triphenylbenzene $C_6H_3(C_6H_5)_3$ at 473 K. From the shape of the angular distribution function $I(S)$ and in particular from the positions of their maxima, we can conclude about the most probable model 1,3,5–triphenylbenzene molecule structure.
- New information on mutual arrangement and orientation of the molecules studied. The most probable binary radial correlation of molecules in liquid 1,3,5–triphenylbenzene was proposed.
- The appearance of clear maxima on the functions of angular-distributions of X-ray radiation indicates the presence of short-range ordering in liquid triphenylbenzene up to the distance of about 20 Å. The function $I(S)$ and DRDF fare sensitive to the shape of molecules of the liquid studied, so they bring information on molecular structures in them and thus also on intra- and intermolecular interactions.
- The applied methods of measurements permitted determination of the mean structural parameters (the inter- and intramolecular distances, the radii of coordination spheres, the coordination numbers) and local ordering of the molecules in the liquid studied.
- X-ray structural analysis of liquid 1,3,5–triphenylbenzene (the theoretical and experimental functions were in good agreement) confirms the model of molecule proposed by Lin and Williams for crystal.
- The proposed method research of 1,3,5–triphenylbenzene in high-temperature can be used in investigation of structure and intermolecular interactions other of liquids.

**Acknowledgement**
This work was supported by Grant NCN Opus Nr. Dec – 2013/09/B/ST4/03711.